\RequirePackage{fix-cm}
\documentclass[pdftex,twocolumn,epjc3]{svjour3}  
\smartqed  
\RequirePackage[T1]{fontenc}
\RequirePackage{graphicx}
 \RequirePackage{mathptmx}      
 \RequirePackage{flushend}
\RequirePackage{latexsym}
\RequirePackage{multirow}
\RequirePackage[numbers,sort&compress]{natbib}

\RequirePackage{siunitx}

\RequirePackage[colorlinks,citecolor=blue,urlcolor=blue,linkcolor=blue]{hyperref}
\usepackage{isotope}
\RequirePackage{lineno}
\RequirePackage{amsmath}
\RequirePackage{etoolbox} 
\RequirePackage{microtype}
\RequirePackage{cancel}
\RequirePackage[normalem]{ulem} 

\newcommand*\linenomathpatch[1]{%
  \cspreto{#1}{\linenomath}%
  \cspreto{#1*}{\linenomath}%
  \csappto{end#1}{\endlinenomath}%
  \csappto{end#1*}{\endlinenomath}%
}

\linenomathpatch{equation}
\linenomathpatch{gather}
\linenomathpatch{multline}
\linenomathpatch{align}
\linenomathpatch{alignat}
\linenomathpatch{flalign}

\journalname{Eur. Phys. J. C}
\begin{document}
\title{Relative Measurement and Extrapolation of the Scintillation Quenching Factor of $\alpha$-Particles in Liquid Argon using DEAP-3600 Data}
\author{
        P.~Adhikari~\thanksref{Carleton}
        \and
        M.~Alp\'{i}zar-Venegas~\thanksref{UNAM}
        \and
        P.-A.~Amaudruz~\thanksref{Triumf}
        \and
        J.~Anstey~\thanksref{Carleton, Mcdonaldinst}
        \and
        D.J.~Auty~\thanksref{Alberta}
        \and
        M.~Batygov~\thanksref{Laurentian}
        \and
        B.~Beltran~\thanksref{Alberta}
        \and
        C.E.~Bina~\thanksref{Alberta, Mcdonaldinst}
        \and
        W.~Bonivento~\thanksref{Cagliari2}
        \and
        M.G.~Boulay~\thanksref{Carleton}
% ...
        (DEAP~Collaboration)\thanksref{email}
}                     
%
%
\thankstext{deceased}{Deceased.}
\thankstext{email}{deap-papers@snolab.ca}
\thankstext{TRPollman}{Currently at Nikhef and the University of Amsterdam, Science Park, 1098XG Amsterdam, Netherlands}

\institute{
    Department  of  Physics,  University  of  Alberta,  Edmonton,  Alberta,  T6G  2R3,  Canada \label{Alberta}
% ...
\and
    Department  of  Physics,  Carleton  University,  Ottawa,  Ontario,  K1S  5B6, Canada \label{Carleton}
% ...
\and
    INFN Cagliari, Cagliari 09042, Italy \label{Cagliari2}
% ...
\and
    School of Natural Sciences, Laurentian University, Sudbury, Ontario, P3E 2C6, Canada \label{Laurentian}
% ...
\and
    Instituto de F\'{i}sica, Universidad Nacional Aut\'{o}noma de M\'{e}xico, A. P. 20-364, Ciudad de M\'{e}xico 01000, Mexico \label{UNAM}
% ...
\and
    TRIUMF, Vancouver, British Columbia, V6T 2A3, Canada \label{Triumf}
% ...
\and
    Arthur B. McDonald Canadian Astroparticle Physics Research Institute, Queen's University, Kingston, ON, K7L 3N6, Canada \label{Mcdonaldinst}
}
%
\date{Received: date / Revised version: \today}
%
\maketitle

\abstract{
The knowledge of scintillation quenching of $\alpha$-particles 
plays a paramount role in understanding $\alpha$-induced backgrounds and 
improving the sensitivity of liquid argon-based direct 
detection of dark matter experiments. 
We performed a relative measurement of scintillation quenching in the MeV energy region
using radioactive isotopes ($^{222}$Rn, $^{218}$Po and 
$^{214}$Po isotopes) present in trace amounts in the DEAP-3600 detector and
quantified the uncertainty of extrapolating the quenching factor 
to the low-energy region.
%
} 

\section{Introduction}
\label{intro}

Over the past decade, liquid argon (LAr) detectors have been extensively used for dark matter direct detection experiments due to the high purity, scalability and excellent scintillation efficiency of this material~\cite{PhysRevD.100.022004,PhysRevD.98.102006,2018EPJP..133..131A,A-Badertscher_2013,RIELAGE2015144}.
These experiments are optimized to primarily measure the scintillation signals induced by low-energy nuclear recoils
that could be produced due to elastic scattering of weakly interacting massive particles (WIMPs), a promising dark matter candidate. 
Because the WIMP interaction rate is extremely low, mitigating background events is an inevitable requirement of such a detection procedure.

The scintillation time profiles of LAr due to electronic recoils (ERs) 
and nuclear recoils (NRs) differ from each other~\cite{BOULAY2006179,Adhikari_2020}. 
This characteristic helps to reduce ER backgrounds due to $^{39}$Ar $\beta$ decays using the pulse-shape discrimination (PSD) technique~\cite{Adhikari_2021}. 
DEAP-3600 achieved the most sensitive limit on the spin-independent WIMP-nucleon cross-section on argon
above 30 GeV/c$^{\rm 2}$ WIMP mass~\cite{PhysRevD.100.022004}.
The sensitivity of the DEAP-3600 experiment is limited by 
the presence of intrinsic $\alpha$-backgrounds. 
The most challenging such backgrounds come from $\alpha$-decays within the trace amount of dust
particulates dispersed in the LAr volume,
and from $\alpha$-decays of $^{210}$Po 
on the flowguides in the neck of the detector, 
following which the scintillation light is shadowed.

In order to correctly model $\alpha$-induced backgrounds, the amount of scintillation light induced by $\alpha$-particles in LAr must be understood over a wide range of energy. This can be quantified by considering the fraction of deposited energy that is dissipated as scintillation photons: this ratio is known as the quenching factor (QF). A number of experimental studies have been performed to measure the scintillation yield due to neutron-induced NRs in LAr in the low-energy region~\cite{PhysRevC.85.065811,Regenfus_2012,PhysRevD.91.092007,PhysRevD.97.112005}.
The scintillation quenching at low energies cannot completely be explained by the Lindhard {\it et al.} theory~\cite{osti_4701226}  
and various models have been proposed to understand the light yield for low-energy NRs in LAr~\cite{PhysRevD.91.092007,PhysRevD.97.112005,MEI200812,instruments5010005}.
In contrast to NRs, few measurements are available for studying scintillation light due to $\alpha$-particles in LAr at high energies using 
a $^{210}$Po source~\cite{DOKE1988291,HITACHI198297,PhysRevB.54.15724}.  
In the absence of a universally accepted $\alpha$-quenching model, we explore the scintillation yield of $\alpha$-particles over a wide energy region (tens of keV to a few MeV) and quantify extrapolation uncertainties.

The main objective of the present work is to generate an 
energy-dependent scintillation quenching curve for $\alpha$-particles in LAr
and probe the uncertainty of extrapolating the $\alpha$-particle QF to the low-energy region for LAr.
As input to the extrapolation procedure, we start from the direct measurement of the
scintillation QF of $\alpha$-particles from $^{210}$Po reported in Ref.~\cite{DOKE1988291},
and perform a relative measurement of the QF using full-energy $\alpha$ peaks from the $^{222}$Rn chain using
DEAP-3600 data.
Using such a relative measurement reduces the impact of systematic uncertainties related to the detector response and absolute energy calibration.
Based on the results in the MeV range, a model is then used to extrapolate the $\alpha$-particle QF into the low-energy region for LAr.

Section~\ref{sec:data} presents a brief 
description of the DEAP-3600 detector as well 
as data selection.
The relative measurement 
of the scintillation QF for $\alpha$-particles
in the MeV energy region is described in 
Section~\ref{sec:measurement}.
In Section~\ref{sec:model}, we extrapolate the energy-dependent $\alpha$-particle QF and estimate the uncertainties down to energy values in the tens of keV range.
Finally, we summarize the results in Section~\ref{sec:discussion}.
\begin{table*}[btph!]
 \caption{The short-lived $\alpha$-decays of interest in LAr~\cite{ENDF-222Rn,ENDF-218Po,ENDF-214Po,web:Lund:LBNL}. The activity of these isotopes in the DEAP-3600 detector was estimated in Ref.~\cite{PhysRevD.100.022004}.}
\label{tab:Table_source_info}
\centering
    \begin{tabular*}{\textwidth}{{@{\extracolsep{\fill}}llllll@{}}}
    \hline\noalign{\smallskip}
      Radioactive & Half-life& Branching &$\alpha$-particle  &Q-value  & Activity in DEAP-3600 \\
      isotope& &ratio (\%) & energy (MeV) & (MeV) & ($\mu$Bq/kg of LAr)  \\
      \hline
       $^{222}$Rn  & 3.8 days& 99.92 &5.489 & 5.590 &0.153 $\pm$ 0.005\\
       $^{218}$Po  & 3.1 min&99.9989 &6.002 & 6.115 &0.159 $\pm$ 0.005\\
       $^{214}$Po  & 164.3 $\mu$s& 99.9895  &7.686  & 7.833 & 0.153 $\pm$ 0.005\\
           \noalign{\smallskip}\hline
    \end{tabular*}
\end{table*}

\section{Detector and data selection}
\label{sec:data}

The DEAP-3600 detector contains
3.3 tonnes of LAr within a spherical acrylic 
vessel (AV) of radius 850 mm. The top 300 mm of 
the spherical detector region contains gaseous argon (GAr).  
A detailed description of the detector is given in 
Ref.~\cite{AMAUDRUZ20191}. The inner 
surface of the AV is coated with a 3~$\mu$m layer of
1,1,4,4-tetraphenyl-1,3-butadiene
(TPB) wavelength shifter which converts 128 nm 
scintillation light produced by LAr to visible light
that peaks at 420 nm.  
The light signal is acquired by 255 inward-facing 
Hamamatsu R5912-HQE photomultiplier tubes (PMTs) which have high quantum efficiency and low radioactivity~\cite{AMAUDRUZ2019373}. The PMT signals are digitized by both high-gain (CAEN V1720) and low-gain (CAEN V1740) waveform digitizer channels. Signals from the 
high-gain digitizer channels are used in this analysis. 

Data are processed using the RAT~\cite{rat} software framework adapted for DEAP-3600 data.
The charge of a pulse detected in a PMT is divided by the average single-PE charge of the PMT, known from calibrations, to obtain the number of recorded photoelectrons (PEs).
This analysis uses the full dataset collected by DEAP-3600 from November 2016 to December 2017,
plus 20\% of the dataset collected from January 2018 to March 2020. This corresponds to 388.4 live-days.
\begin{figure}[btph!]
    \centering
    \includegraphics[width=\columnwidth]{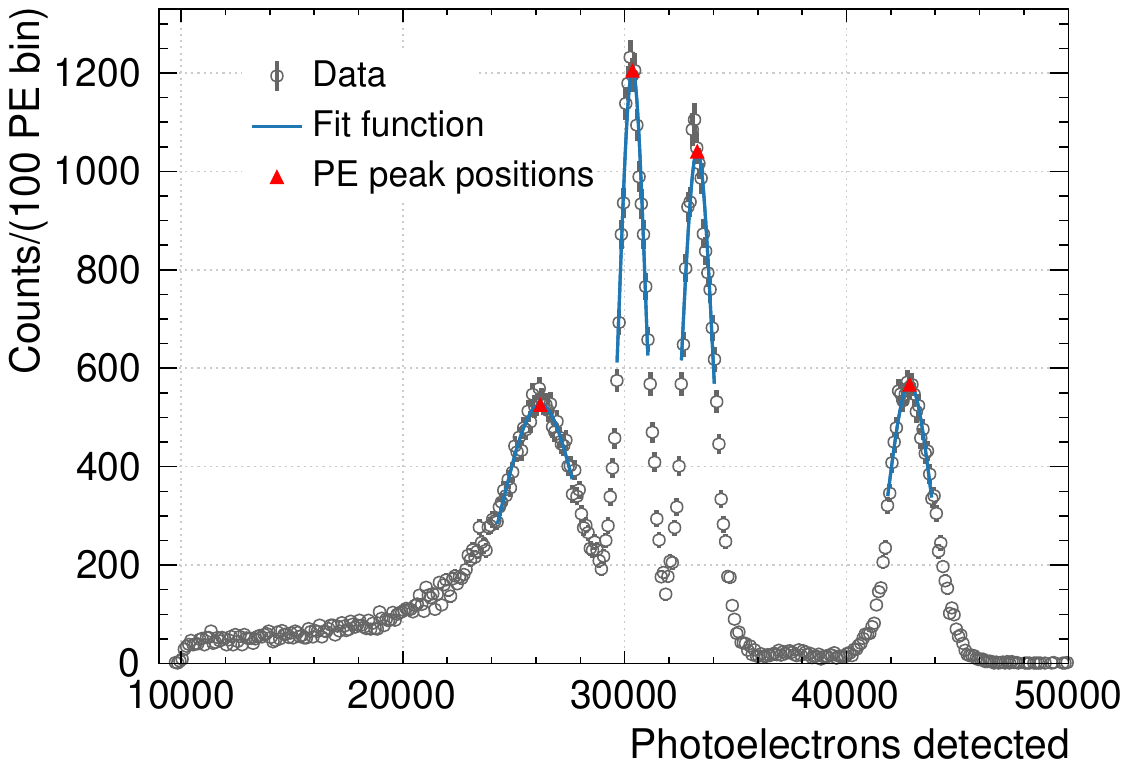}
    \includegraphics[width=\columnwidth]{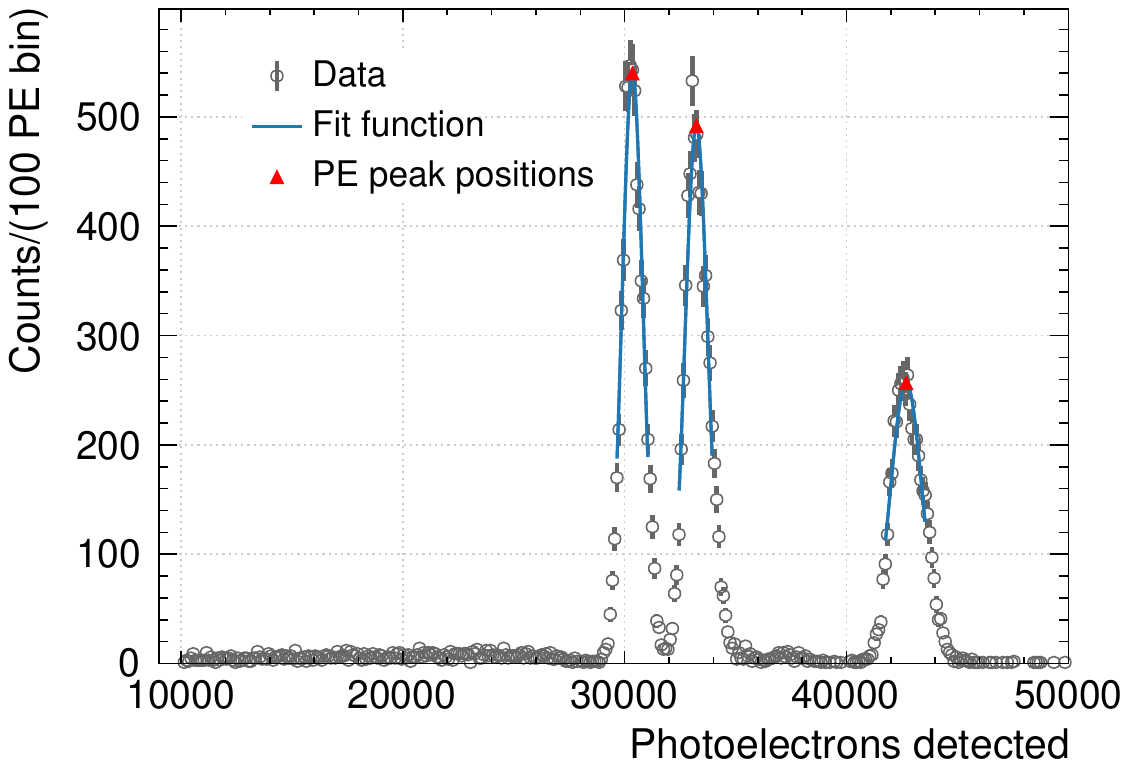}
    \caption{Distribution of the number of PE 
    detected from $\alpha$-decay events the position of which is reconstructed within radius \mbox{0--850~mm} (top histogram) and \mbox{0--600~mm} (bottom histogram) from the origin of the detector. From left to right, the peaks are from the decay of the $^{210}$Po, $^{222}$Rn, $^{218}$Po and $^{214}$Po isotopes. Each peak is fitted by a Gaussian distribution (blue line).  
    Red solid points are the detected peak PE positions  from each Gaussian fit.}
    \label{fig:bulkalphaPE}
\end{figure}

Events are observed due to $\alpha$-decays from the $^{222}$Rn 
isotope and its descendants, occurring inside the detector including the LAr target~\cite{PhysRevD.100.022004}.
Starting from $^{222}$Rn, three $\alpha$-particles and two $\beta$-particles are emitted following the sequence of transitions before producing the long-lived isotope $^{210}$Pb, the chain with the highest probability being:
\begin{equation}
   \rm  \isotope[222][86]{Rn}\xrightarrow[]{\alpha}
   \isotope[218][84]Po\xrightarrow[]{\alpha}
   \isotope[214][82]Pb\xrightarrow[]{\beta}
   \isotope[214][83]Bi\xrightarrow[]{\beta}
   \isotope[214][84]Po\xrightarrow[]{\alpha}
   \isotope[210][82]Pb. \nonumber
\end{equation}
In this work, we consider
$\alpha$-particles emitted from short-lived decays and depositing
their full energy in the LAr volume.
These $\alpha$-emitting isotopes are detailed in Table~\ref{tab:Table_source_info}.
From the long-lived $^{210}$Pb isotope (22.2 year half-life),
the $^{210}$Po isotope is produced
following two $\beta$-decays:
\begin{equation}
 \rm   \isotope[210][82]Pb\xrightarrow[]{\beta}
   \isotope[210][83]Bi\xrightarrow[]{\beta}
   \isotope[210][84]Po\quad. \nonumber
\end{equation}
We do not consider $\alpha$-decays from 
$^{210}$Po in the relative QF measurement,
because in DEAP-3600 these decays mostly originate from 
the inner surface of the AV:
the resulting $\alpha$-particles deposit
an unknown portion of their energy into the acrylic and TPB 
before creating scintillation signals 
in LAr~\cite{PhysRevD.100.022004}. 

Event selection cuts are applied to the data in order to select $\alpha$-decay candidates.
First, the same run selection, trigger requirements and data cleaning cuts as in Ref.~\cite{PhysRevD.100.022004} are applied.
Then, to remove pre-trigger pileup and post-trigger pileup from coincidence events, the event trigger time must be within the range [2250, 2700]~ns in the trigger window, be recorded at least 20~$\mu$s after the previous event's trigger time, and with fewer than 4 pulses recorded by PMTs in the first 1600~ns of the event.
Events are rejected if more than 75\% of the PE in the event were detected by a single PMT.
To select events from the NR band and reject ERs, a PSD cut is applied with the requirement that at least 55\% of the PE are detected within [-28, 150]~ns of the trigger time.
Finally, the reconstructed position of the event must be within the LAr volume.

Fig.~\ref{fig:bulkalphaPE} shows the PE distribution of the selected events at high PE,
with $\alpha$ peaks observed from the decay of $^{210}$Po, $^{222}$Rn, $^{218}$Po and $^{214}$Po isotopes.  Each peak is fitted by a
Gaussian distribution using TMinuit~\cite{minuit} in ROOT~\cite{BRUN199781} and the mean of the each fit function is considered as the PE peak value.

\section{Relative measurement of the quenching factor} \label{sec:measurement}

\subsection{Method}
The scintillation quenching, QF$_{\alpha}$,  of $\alpha$-particles with energy 
E$_{\alpha}$ can be expressed 
by the following equation:
\begin{equation}
{\rm QF}_{\alpha}=\frac{{\rm PE}_{\alpha}}{Y \times E_{\alpha, {\rm dep}}} \quad,
\label{Eq:Eq_QF}
\end{equation}
where ${\rm PE}_{\alpha}$ is the detected number of PE at the peak, $E_{\alpha, {\rm dep}}$ is the total 
energy deposited by the $\alpha$-particles within LAr, and 
$Y$ (in PE/keV) is the scintillation light yield measured for $\gamma$-rays,
which is assumed to be linear over the 5~to~8~MeV range.  
For the $\alpha$-decays observed on-peak, the full energy is
deposited within the LAr volume of the detector, therefore
we take $E_{\alpha, {\rm dep}} = E_{\alpha}$ from Table~\ref{tab:Table_source_info}.
We assume that the number of detected PE generated by the heavy nuclear recoils in these events is negligible compared to the amount generated by the full-energy $\alpha$ particles.

The DEAP-3600 detector is calibrated at low energy from around 
565 keV to 1.3 MeV  using $\beta$-decays 
from naturally present $^{39}$Ar isotopes in LAr and an
external $^{22}$Na $\gamma$ source. The energy response 
function is linear over a wide range of energies, 
whereas a non-linear response has been observed above 
a few MeV because of the saturation of detected pulses. 
Saturation of the PMTs and digitizers can occur; digitizer clipping effects start at lower PE compared to the region where PMT non-linearity effects are observed. 
In energy reconstruction, a correction algorithm is used to deal with such effects.
As a first step, this algorithm identifies clipped pulses.
For each such pulse, the charge measured in high-gain digitizer channels is corrected using a function that was obtained by comparing 
the pulse integrals of 
a large number of unsaturated  
pulses recorded by high-gain digitizer channels to the corresponding pulses recorded by low-gain digitizer channels. 
This reconstructed charge is used to recalculate the number of PE, after which another correction is applied to account for the non-linear PMT response which appears due to space-charge effects within the PMT dynode structure.  
For this purpose, a sigmoidal function is used to model charge growth in a PMT in the presence of space-charge. This function controls both the PE value at which saturation begins and the nature of the gradual change in the saturation effect through three parameters determined by tuning with data.
Pulses from events originating close to the AV surface are more likely to be saturated. The magnitude of the entire correction procedure is typically less than 1\%; it can range up to 10\% for most pulses, and rarely up to 50\% for pulses from high-energy events closest to the AV surface.
This has an indirect dependence on the reconstructed position of the events, systematic uncertainties from which are taken into account in the next section. A detailed description of this procedure is provided in Ref.~\cite{JM_PLSC}.

This analysis uses three data points:
the measurement of QF$_{\alpha, \rm ^{210}Po} = 0.710 \pm 0.028$ at 5.305 MeV by T.~Doke {\it et~al.}~\cite{DOKE1988291};
and the following two relative measurements from DEAP-3600.
Based on Eq.~\ref{Eq:Eq_QF}, we define the ratios $R_2$ and $R_3$ as follows:

\begin{equation}
\frac{{\rm QF}_{\alpha, \rm ^{218}Po}}{{\rm QF}_{\alpha,  \rm ^{222}Rn} } =\frac{{\rm PE}_{\alpha, \rm ^{218}Po}}{{\rm PE}_{\alpha, \rm ^{222}Rn}}\times \frac{E_{\alpha, \rm ^{222}Rn}}{E_{\alpha, \rm ^{218}Po}}
\equiv R_{2}\times \frac{E_{\alpha,1}}{E_{\alpha,2}}, \label{Eq:QF2R2}
\end{equation}
\begin{equation}
\frac{{\rm QF}_{\alpha, \rm ^{214}Po}}{{\rm QF}_{\alpha, \rm ^{222}Rn}} =\frac{{\rm PE}_{\alpha, \rm ^{214}Po}}{{\rm PE}_{\alpha, \rm ^{222}Rn}}\times \frac{E_{\alpha, \rm ^{222}Rn}}{E_{\alpha, \rm ^{214}Po}}
\equiv R_{3}\times \frac{E_{\alpha,1}}{E_{\alpha,3}}\label{Eq:QF3R3}.
\end{equation}
where on the right, the subscripts 1, 2, 3 are used for $^{222}$Rn, $^{218}$Po and $^{214}$Po respectively.

By taking the ratios of the peak PE values for $\alpha$-particles within the 5~to~8~MeV range, the effect of non-linearities in the light detection efficiency on  
the estimation of the QF is reduced, and the analysis 
becomes less sensitive to the absolute energy calibration at high energy.
These ratios for $^{218}$Po and $^{214}$Po relative to $^{222}$Rn are calculated using Gaussian fits to the whole dataset as shown in Fig.~\ref{fig:bulkalphaPE} (top):
the resulting measured values of $R_2$ and $R_3$ are given in Table~\ref{tab:QF_table}.

\subsection{Systematic uncertainties}
\label{subsec:uncertainty}

The light yield $Y$ depends on various 
detector parameters, such as
the photo-detection efficiency of the PMTs,
the reflectivity of the AV surface, 
the roughness of the AV inner surface, 
the efficiency and thickness of TPB, etc. 
Some of these properties could vary
throughout the data taking period or different 
locations within the detector.
Therefore, the uncertainties on $R_2$ and $R_3$ 
are estimated in terms of their observed variations
as a function of reconstructed position and time of 
occurrence of the $\alpha$-decays within the LAr
throughout the data-taking period.  

To estimate uncertainties due to $\alpha$-decay position ($\sigma_p$), 
the detector volume is divided into four concentric 
spherical regions where the distance to the origin of the detector is
\mbox{0--200}~mm, 200--400~mm, 400--600~mm and 600--800~mm. 
The LAr volume which is within 50~mm from the inner surface 
of the AV is not considered here because the PE peak from the $\alpha$-decay of $^{210}$Po on the AV surface is the dominant contribution there. 
For each region, the peak PE values from $^{222}$Rn, $^{218}$Po and $^{214}$Po are each determined as the mean of the corresponding Gaussian fit.
Averaged over all four regions, the ratios of $^{218}$Po and $^{214}$Po peak PE to the $^{222}$Rn peak PE are respectively  $1.096 \pm 0.001$ and $1.408 \pm 0.005$.

In order to estimate the uncertainties due to time of occurrence of $\alpha$-decays in the detector ($\sigma_t$), the entire dataset is divided into 
twenty-one time bins, each covering sixty days except the last bin. The peak PE ratios are estimated for each time bin.  Unlike the previous case, events originating near the inner surface of the detector are also considered here. 
Averaged over time bins, the peak PE ratios relative to $^{222}$Rn are $1.096 \pm 0.002$ for $^{218}$Po and $1.412 \pm 0.003$ for $^{214}$Po.

Assuming the systematic uncertainties from event position and time are uncorrelated, their sum in quadrature is taken as the absolute uncertainty $\sigma_i = \sqrt{(\sigma_{p,i})^2 + (\sigma_{t,i})^2}$ on the peak PE ratio $R_i$, overall resulting in 
$R_2 = 1.096 \pm 0.002$ and 
$R_3 = 1.411 \pm 0.006$.

If we further assume that at 5.489~MeV, ${\rm QF}_{\alpha, \rm ^{222}Rn} = {\rm QF}_{\alpha, \rm ^{210}Po}$ measured in Ref.~\cite{DOKE1988291}, it follows from rearranging Eq.~\ref{Eq:QF2R2}~and~\ref{Eq:QF3R3} that
${\rm QF}_{\alpha, \rm ^{218}Po} = 0.712 \pm 0.001$ and
${\rm QF}_{\alpha, \rm ^{214}Po} = 0.716 \pm 0.003$ 
with the uncertainties due to relative PE peak positions only.
The total uncertainty on these QF values is dominated by the absolute uncertainty of $\pm 0.028$ from the measurement of ${\rm QF}_{\alpha, \rm ^{210}Po}$.

\begin{table*} [tbh]
 \caption{ Quenching factor of $\alpha$-particles obtained from the relative measurement using $^{222}$Rn, $^{218}$Po and $^{214}$Po decays within the DEAP-3600 detector.  
    The measured value by Ref.~\cite{DOKE1988291} of the QF for $\alpha$-particles from $^{210}$Po decays is also shown.}   
    \label{tab:QF_table}
    \centering
    \begin{tabular*}{\textwidth}{{@{\extracolsep{\fill}}lllllll@{}}}
    \hline\noalign{\smallskip}
Radioactive &Energy of & Ratio of PE peak & Uncertainty on  & Quenching & Uncertainty on  &  Absolute \\
isotope & $\alpha$-particle  & to $^{222}$Rn PE peak  & the peak PE ratio & factor &  QF$_\alpha$ due to  & uncertainty  \\
  & (MeV)  & ($R_i$) & ($\sigma_i$) & (QF$_\alpha$) &  PE peak ratios & on QF$_\alpha$    \\

 \hline
$^{210}$Po  & 5.305 & - & - & 0.710~\cite{DOKE1988291}&- &0.028~\cite{DOKE1988291}\\   
 
$^{218}$Po & 6.002& 1.096 & 0.002 & 0.712 & 0.001 & -\\
$^{214}$Po & 7.686 & 1.411 & 0.006 & 0.716& 0.003 & -\\
\noalign{\smallskip}\hline
    \end{tabular*}
   
\end{table*}

\section{Extrapolation of the quenching factor}
\label{sec:model}

With the help of the existing experimental data~\cite{PhysRevD.72.072006,ARNEODO2000147,AKIMOV2002245,BRUNETTI2005265}, Mei {\it et al.}~\cite{MEI200812} proposed a prescription which 
predicts that the scintillation 
quenching of neutron-induced nuclear 
recoils in noble liquids from a few tens of keV to a few hundreds of keV
is influenced by two different mechanisms:
reduction of energy transferred to electrons, and consequently production of a smaller number of excimers and ions; and 
reduction of scintillation yield due to high ionization 
and excitation density.

Here, we use a model that considers two independent quenching effects:
a ``nuclear QF'' accounting for the fraction of energy lost 
by a recoiling nucleus
as a result of nuclear collisions, and
an ``electronic QF'' 
accounting for non-radiative de-excitation of excimers produced by energy transferred to atomic electrons.
The product of these two factors will be taken as the final result.

\begin{figure}[tbh]
    \centering
    \includegraphics[width=\columnwidth]{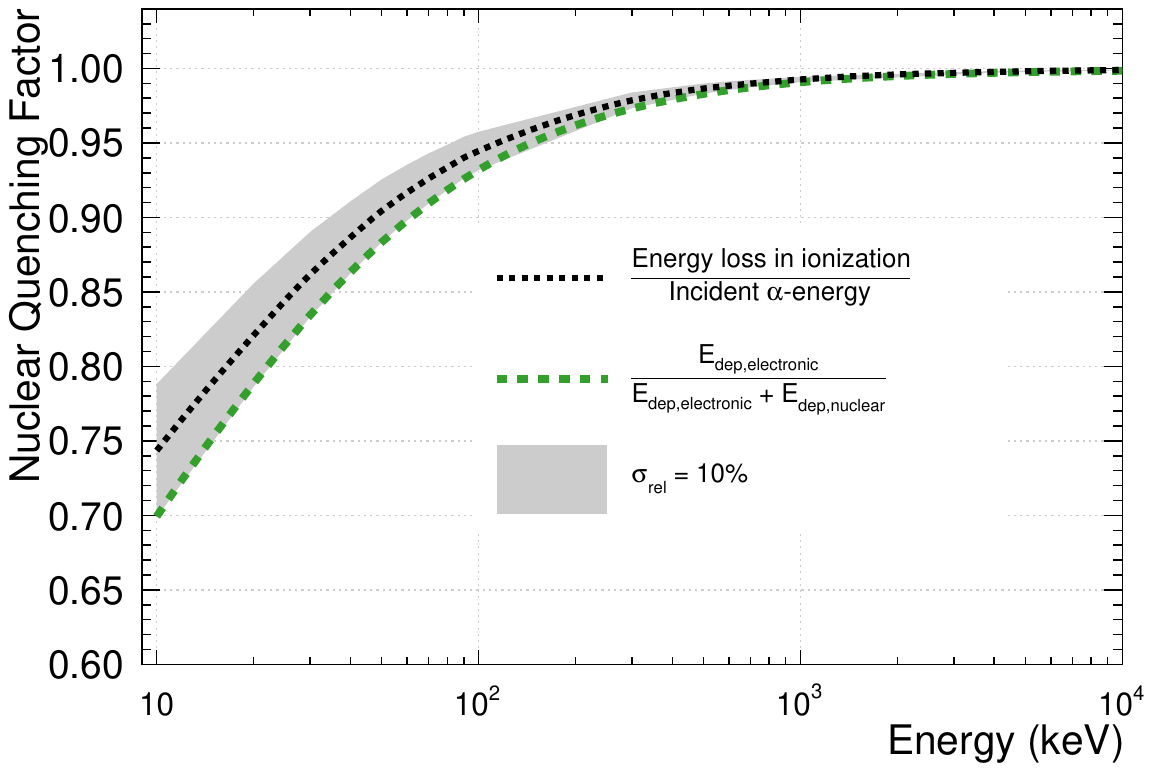}
    \caption{ Nuclear QF curve for $\alpha$-particles
    estimated using SRIM stopping power tables and Eq.~\ref{eq:Lind-styleEq} (green)
    and TRIM simulation (black).}
    \label{fig:LindQF}
\end{figure}

\subsection{Nuclear quenching factor}
\label{subsec:nuclearQF}

The ratio of energy transferred to electrons
($E_{\rm dep, elec}$)
to the total energy deposition, where this denominator includes the part responsible for nuclear translation motion ($E_{\rm dep, nucl}$) along the $\alpha$-particle's track,
is defined as the nuclear QF:
\begin{equation}
    {\rm QF}_{\alpha}^{\rm nucl}=\frac{E_{\rm dep, elec}}{E_{\rm dep, elec} + E_{\rm dep, nucl}} \quad,
    \label{eq:Lind-styleEq}
\end{equation}
and the uncertainty on this calculation is taken as

\begin{align}
    \Delta{\rm QF}_{\alpha}^{\rm nucl}
    &=\left[\left(\frac{\partial {\rm QF}_{\alpha}^{\rm nucl}}{\partial E_{\rm dep, elec}} \sigma_{\rm elec} \right)^2 
    + \left(\frac{\partial {\rm QF}_{\alpha}^{\rm nucl}}{\partial E_{\rm dep, nucl}} \sigma_{\rm nucl} \right)^2 \right.\\\nonumber
    &+ \left. 2\rho\sigma_{\rm elec} \sigma_{\rm nucl}\left(\frac{\partial {\rm QF}_{\alpha}^{\rm nucl}}{\partial E_{\rm dep, elec}}\right)\left(\frac{\partial {\rm QF}_{\alpha}^{\rm nucl}}{\partial E_{\rm dep, nucl}} \right)  \right]^{1/2}\\\nonumber
    &= \left[ \left(\frac{E_{\rm dep, nucl}\cdot\sigma_{\rm elec}}{\left(E_{\rm dep, elec} + E_{\rm dep, nucl}\right)^{2}}\right)^{2} \right. \\\nonumber
    &+ \left(\frac{-E_{\rm dep, elec}\cdot\sigma_{\rm nucl}}{\left(E_{\rm dep, elec} + E_{\rm dep, nucl}\right)^{2}}\right)^{2} \\
    &- \left.2\rho\cdot\frac{E_{\rm dep, nucl}\cdot E_{\rm dep, elec}\cdot\sigma_{\rm elec} \cdot\sigma_{\rm nucl}}{\left(E_{\rm dep, elec} + E_{\rm dep, nucl}\right)^{4}}\right]^{1/2}\\
    &=\sqrt{2(1-\rho)}\frac{\ E_{\rm dep, elec}\ E_{\rm dep, nucl}}{(E_{\rm dep, elec} + E_{\rm dep, nucl})^2} \sigma_{\rm rel} \quad
    \label{eq:Lind-styleEqUnc}
\end{align}

where 
$\rho = -1$ is the correlation coefficient between $E_{\rm dep, elec}$ and $E_{\rm dep, nucl}$ which are anti-correlated for a fixed total energy deposition. In the last step it is assumed that the relative uncertainties are similar :$\frac{\sigma_{\rm elec}}{E_{\rm dep, elec}}=\frac{\sigma_{\rm nucl}}{E_{\rm dep, nucl}} = \sigma_{\rm rel}$.

To calculate this nuclear QF as a function of energy, 
TRIM (TRansport of Ions in Matter) simulations~\cite{SRIM-TRIM} are performed for $\alpha$-particle energies between 10~keV and 10~MeV. 
This simulation provides the 
energy loss in ionization, phonon generation and radiation damage.  The option named ``Ion Distribution and Quick Calculation of Damage'' is used to minimize the detailed estimation of target damage. To calculate the nuclear QF, the energy loss  
from ionization estimated using TRIM simulations is divided by the incident energy of the $\alpha$-particle. 
The results are shown in Fig.~\ref{fig:LindQF}.

\begin{figure} [h!tpb]
    \centering
    \includegraphics[width=\columnwidth]{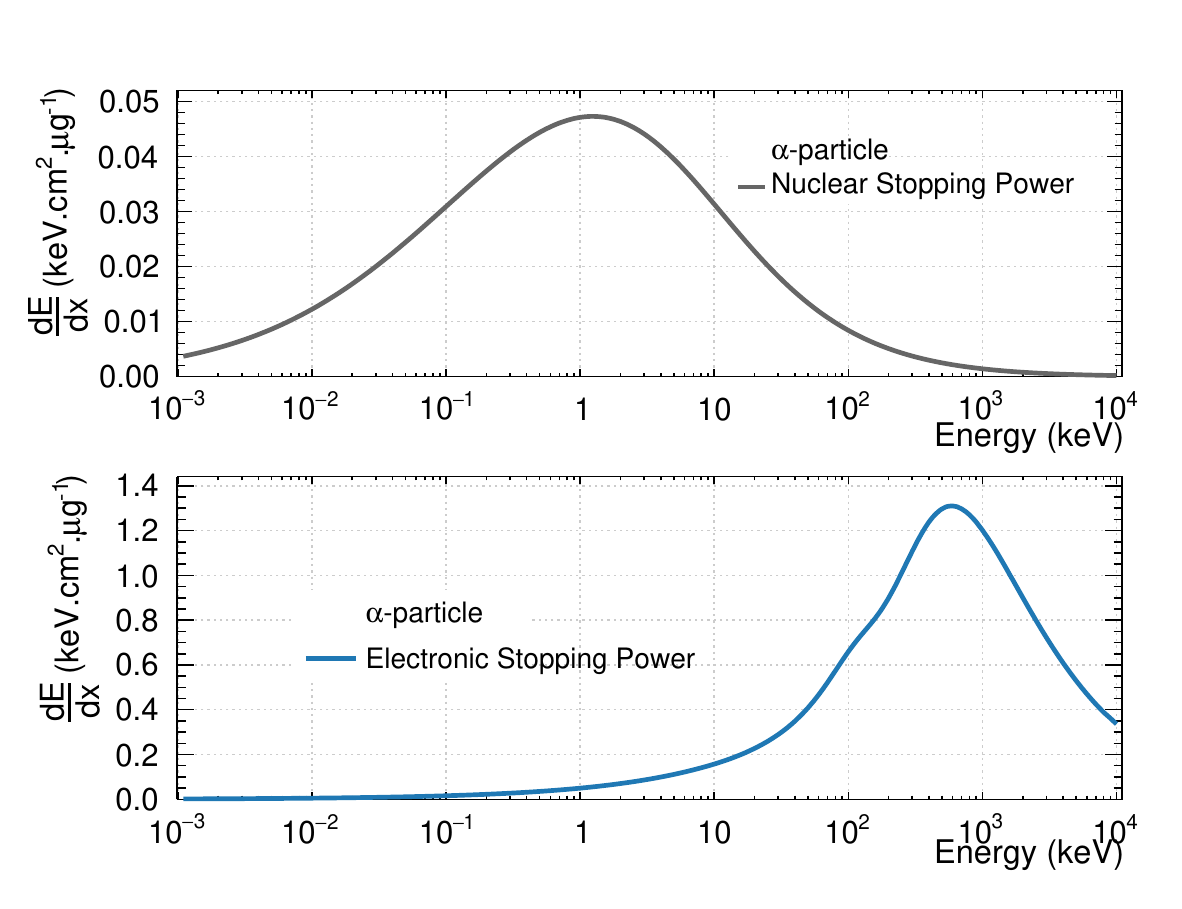}
    \caption{Stopping power curves for $\alpha$-particles in LAr from SRIM-2013~\cite{Ziegler1985}.}
    \label{fig:stopping_power_compare}
\end{figure}

Also shown in Fig.~\ref{fig:LindQF} is the result of a cross-check calculation of the nuclear QF where the electronic and nuclear stopping power values are taken from SRIM (The Stopping and Range of Ions in Matter)-2013~\cite{Ziegler1985}
as shown in Fig.~\ref{fig:stopping_power_compare}. 
The deposited energy 
is calculated in steps of 1~nm along the path of the $\alpha$-particle track.
SRIM calculates the separate electronic and nuclear stopping powers of the $\alpha$ particle; this accounts for the amount of energy transferred to atomic electrons and nuclei directly, but does not account for the subsequent scatters and the final state distribution of the energy deposited.
The resulting nuclear QF curve for $\alpha$-particles 
in LAr is calculated using Eq.~\ref{eq:Lind-styleEq}. 

In the present study, the values of the nuclear QF obtained using TRIM simulations are used in the combined QF calculation of Section~\ref{subsec:comb_results}.
The nuclear QF values from TRIM are about 2\% to 6\% higher than the values obtained using the SRIM stopping power in the range 10--60 keV.  
This difference decreases with increasing $\alpha$ energy
and becomes less than 1\% from around 300 keV. 
The stopping power calculation from SRIM deviates from the experimental data points by a maximum  of approximately 10\% for helium ions  within an argon gas target~\cite{SRIM-TRIM}.
Therefore, the relative uncertainty $\sigma_{\rm rel}$ is considered as 10\% here. The nuclear QF uncertainty is calculated  using Eq.~\ref{eq:Lind-styleEqUnc}
and shown in gray colored band in Fig.~\ref{fig:LindQF}.

\subsection{Electronic quenching factor}
\label{subsec:BirksQF}

Next, using Birks's formalism~\cite{JBBirks_1951},
we account for the non-radiative de-excitation of excimers by interactions with atomic electrons.
For the fraction of the energy deposited in a small step that is converted to scintillation light, we write:
\begin{equation}
    \frac{dy}{dE} = \frac{A}{1+B\frac{dE}{dx}} \quad,
    \label{eq:LY_diff}
\end{equation}
where $\frac{dE}{dx}$ is the electronic stopping power at each step along the $\alpha$-particle's track,
and $A$ and $B$ are constants for any $\alpha$-particle energy.
In this model, $A$ accounts for average quenching due to the density of the deposition within the track core,
and $B$ accounts for the change in density with energy.

The total energy that goes into scintillation along the $\alpha$-particle track is then:
\begin{equation}
    y(E_{\alpha}) = A \int_{0}^{E_{\alpha}}\frac{dE}{1+B\frac{dE}{dx}} \quad.
    \label{eq:LY_int}
\end{equation}
 We define the electronic quenching factor from this model as the fraction of energy that goes into scintillation:  
\begin{equation}
{{\rm QF}_{\alpha}^{M}} =\frac{y(E_{\alpha})}{E_{\alpha}}=\frac{A}{E_{\alpha}}\int_{0}^{E_\alpha}\frac{dE}{1+B\frac{dE}{dx}} \quad.
\label{eq:BirksEq}
\end{equation}

The model predictions for the ratios $R_i$ are likewise:
\begin{equation}
     R_{i}^{M} = \frac{y(E_{\alpha,i})}{y(E_{\alpha,1})}
     = \frac{\int_{0}^{E_{\alpha,i}}\frac{dE}{1+B\frac{dE}{dx}} }
     {\int_{0}^{E_{\alpha,1}}\frac{dE}{1+B\frac{dE}{dx}}} \quad.
\end{equation}

\begin{figure}[tbp]
    \centering
    \includegraphics[width=\columnwidth]{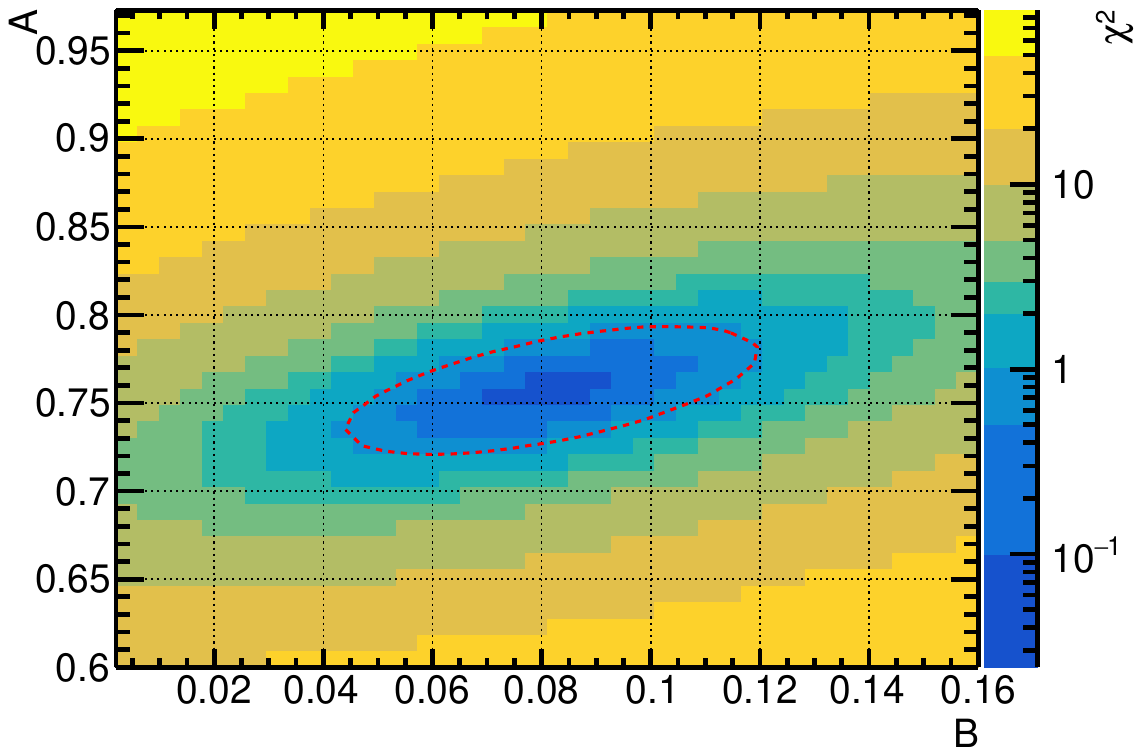}
    \caption{$\chi^{2}$ from Eq.~\ref{Eq:chi2_eq} shown as a function of $A$ and $B$. The red-dotted line represents the 1$\sigma$ contour drawn in the $(A,B)$ parameter space.}
    \label{fig:A_B-chi2}
\end{figure}

\begin{figure}[tbp]
    \centering
     \includegraphics[width=\columnwidth]{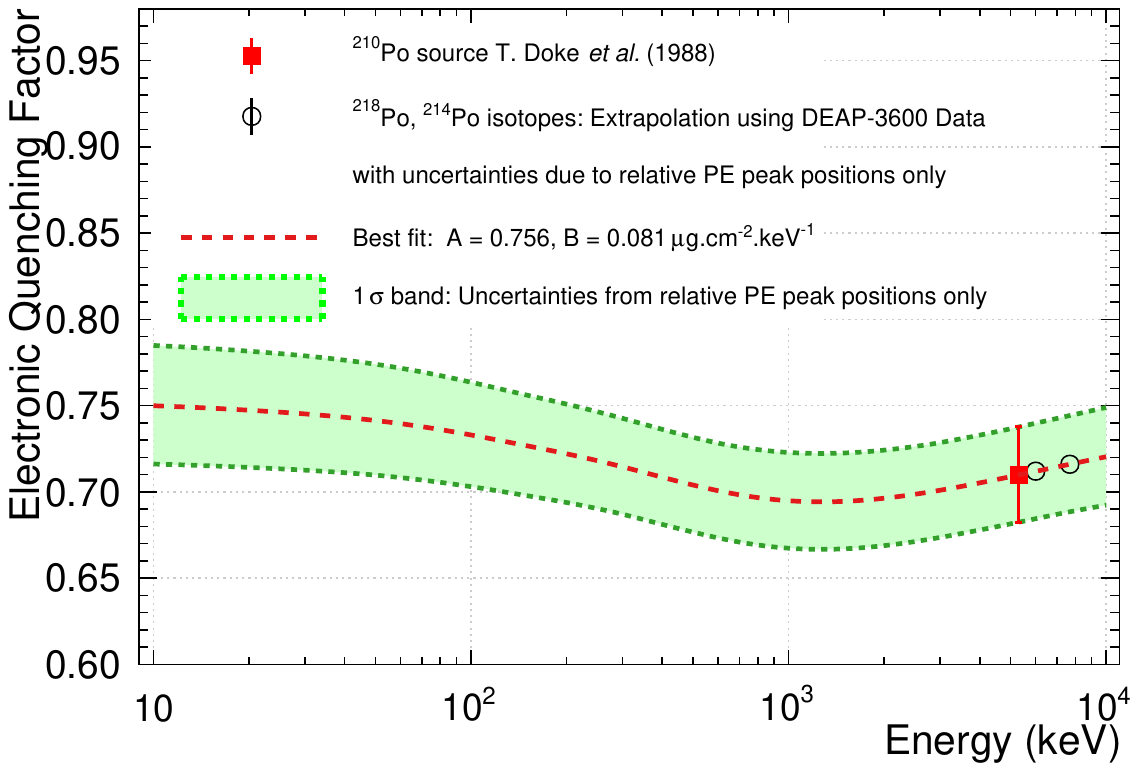}
    \caption{ Electronic QF as a function of $\alpha$-particle energy. The best-fit QF curve is shown with the red dashed line. The green shaded region is the 1$\sigma$ band considering  only the uncertainties of the relative measurement, as a function of energy. This band encompasses  the absolute uncertainty from the measurement of ${\rm QF}_{\alpha, \rm ^{210}Po}$ which is the dominant uncertainty in this analysis. The red solid square represents T. Doke's measurement of the scintillation quenching of $\alpha$-particles emitted from the decay of $^{210}$Po~\cite{DOKE1988291}.   Two black open circles display the DEAP-3600 relative measurements from $^{218}$Po and $^{214}$Po (see Table~\ref{tab:QF_table}). 
    }
    \label{fig:BirksFit}
\end{figure}

We constrain the values of the parameters $A$ and $B$ using the measured values of QF$_{\alpha, \rm ^{210}Po}$ from Ref.~\cite{DOKE1988291} and the two $R_i$ from Section~\ref{sec:measurement}, by minimizing
\begin{equation}
    \chi^{2} = \frac{({\rm QF}_{\alpha,\rm ^{210}Po} - {\rm QF}_{\alpha,\rm ^{210}Po}^{M})^{2}}{\sigma_{0}^{2}}
    + \sum_{i=2}^{3} \frac{\left(R_{i}-R_{i}^{M}\right)^{2}}{\sigma_{i}^{2}}
    \label{Eq:chi2_eq}
\end{equation}
where as reported above the uncertainties are $\sigma_{0} = 0.028$, $\sigma_{2} = 0.002$ and $\sigma_{3} = 0.006$.
Using the electronic stopping power values from SRIM-2013~\cite{Ziegler1985}, this $\chi^{2}$ is calculated in a grid of $(A,B)$ values over the range $A = (0.600-1.000)$ and $B = (0.001 - 0.160)$: the result is shown in Fig.~\ref{fig:A_B-chi2}.
The $\chi^{2}$ is observed to be smooth, with a minimum at $A = 0.756$ and $B = 0.081\ {\rm \mu g\cdot cm^{-2}\cdot keV^{-1}}$.
The resulting best-fit electronic QF curve is shown in Fig.~\ref{fig:BirksFit}.

Based on the $\chi^2$ function, a 1$\sigma$ contour is drawn in $(A, B)$ parameter space where on the contour $\chi^2 = \chi^2_{\rm min} + 1$, with $\chi^2_{\rm min}$ the minimum $\chi^2$ value (corresponding to the best fit): 
this contour is also shown in Fig.~\ref{fig:A_B-chi2}. 
For different $(A, B)$ combinations along this contour, the electronic QF is calculated as a function of $\alpha$-particle energy using Eq.~\ref{eq:BirksEq}.
The 1$\sigma$ bands in Fig.~\ref{fig:BirksFit} are generated by using the QF curves from the worst fits that are consistent with the data at the 1$\sigma$ level.

\subsection{Combined results}
\label{subsec:comb_results}

The final energy-dependent QF curve is obtained by taking the product of the best-fit electronic QF curve and the nuclear QF curve from TRIM.
According to these results, at high energy the electronic quenching mechanism is dominant,
while at low energy nuclear quenching becomes important.

\begin{figure}[h!tpb]
    \centering
    \vspace{-1em}
    \includegraphics[width=\columnwidth]{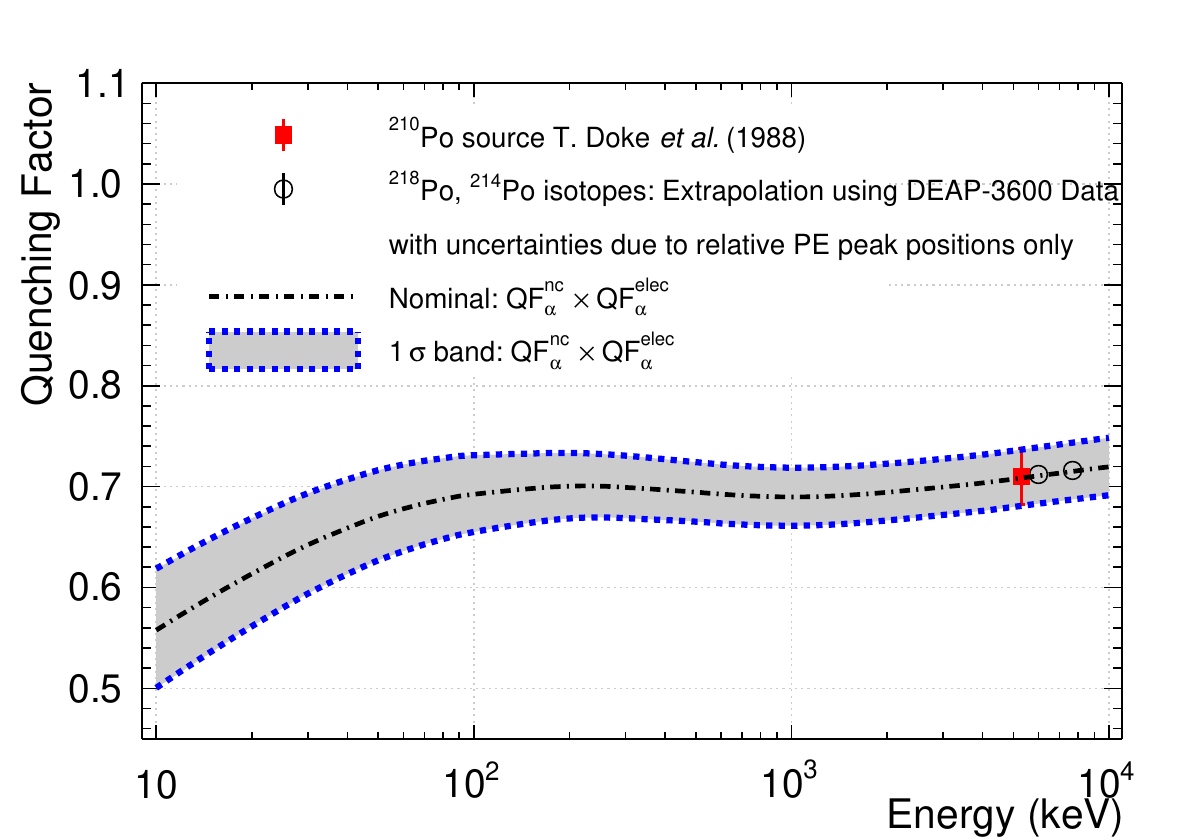}
    \caption{     Energy-dependent 
    scintillation QF curve for $\alpha$-particles within LAr, as a function of their energy on a logarithmic scale. 
    The nominal and $\pm 1\sigma$ QF curves are
    the product of the electronic QF (see Fig.~\ref{fig:BirksFit}) with the nuclear QF from TRIM (see Fig.~\ref{fig:LindQF}).
    }
    \label{fig:FinalQF}
\end{figure}

Fig.~\ref{fig:FinalQF} displays the scintillation QF curve for $\alpha$-particles in LAr and the corresponding $\pm 1 \sigma$ uncertainty band over the energy range 10~keV--10~MeV, from the combination of the electronic QF with the nuclear QF.
This uncertainty band comes from the combination of the $1\sigma$ band on the electronic QF shown in Fig.~\ref{fig:BirksFit} with the nuclear QF uncertainty calculated using Eq.~\ref{eq:Lind-styleEqUnc} taking 
$\sigma_{\rm rel} = 10\%$.

Relative light yield  measurements were performed by A. Hitachi {\it et~al.}~\cite{HITACHI198297} within liquid argon  by measuring the pulse heights of $\alpha$-particles from  
  $^{210}$Po, $^{212}$Bi, $^{252}$Cf and $^{212}$Po relative to 6 MeV $\alpha$-particles.  
The resulting estimates of the quenching factors are consistent with our results within uncertainties.

\section{Summary and outlook} \label{sec:discussion}

In summary, $\alpha$-particles from radioactive decays detected in LAr with the DEAP-3600 detector 
are used to perform a relative measurement of the QF at energies between 5.489 and 7.686 MeV, corresponding to the full-energy $\alpha$ peaks.  
One advantage of the relative measurement procedure is to reduce the impact of light-yield non-linearity effects observed in DEAP-3600 data above a few MeV.
We extrapolated the QF values into the 
low-energy region down to 10 keV, and also assigned uncertainties to this extrapolation. The energy-dependent QF curve is utilized in the analysis of backgrounds to the WIMP search. Measurements of the QF at various energies are currently underway in order to validate these results.

\begin{acknowledgement}
We thank the Natural Sciences and Engineering Research Council of Canada (NSERC),
the Canada Foundation for Innovation (CFI),
the Ontario Ministry of Research and Innovation (MRI), 
and Alberta Advanced Education and Technology (ASRIP),
the University of Alberta,
Carleton University, 
Queen's University,
the Canada First Research Excellence Fund through the Arthur B.~McDonald Canadian Astroparticle Physics Research Institute,
Consejo Nacional de Ciencia y Tecnolog\'ia Project No. CONACYT CB-2017-2018/A1-S-8960, 
DGAPA UNAM Grants No. PAPIIT IN108020 and IN105923, 
and Fundaci\'on Marcos Moshinsky,
the European Research Council Project (ERC StG 279980),
the UK Science and Technology Facilities Council (STFC) (ST/K002570/1 and ST/R002908/1),
the Leverhulme Trust (ECF-20130496),
the Russian Science Foundation (Grant No. 21-72-10065),
the Spanish Ministry of Science and Innovation (PID2019-109374GB-I00) and the Community of Madrid (2018-T2/ TIC-10494), 
the International Research Agenda Programme AstroCeNT (MAB/2018/7)
funded by the Foundation for Polish Science (FNP) from the European Regional Development Fund,
and the Polish National Science Centre (2022/47/B/ST2/02015).
Studentship support from
the Rutherford Appleton Laboratory Particle Physics Division,
STFC and SEPNet PhD is acknowledged.
We thank SNOLAB and its staff for support through underground space, logistical, and technical services.
SNOLAB operations are supported by the CFI
and Province of Ontario MRI,
with underground access provided by Vale at the Creighton mine site.
We thank Vale for their continuing support, including the work of shipping the acrylic vessel underground.
We gratefully acknowledge the support of the Digital Research Alliance of Canada,
Calcul Qu\'ebec,
the Centre for Advanced Computing at Queen's University,
and the Computational Centre for Particle and Astrophysics (C2PAP) at the Leibniz Supercomputer Centre (LRZ)
for providing the computing resources required to undertake this work.
\end{acknowledgement}

%
%
\bibliographystyle{spphys}
\bibliography{main.bib}
%
%

\end{document}